\documentclass[sigconf]{acmart}
\usepackage{booktabs}

\setcopyright{none}

\acmConference[WWW'18]{ACM WWW conference}{April 2018}{Lyon, France} 
\acmYear{2018}

\usepackage{color}
\usepackage{soul}
\usepackage{url}
\usepackage{multirow}
\usepackage{lipsum}                    
\usepackage{xargs}                      
\usepackage{todonotes}
\usepackage{adjustbox}
\usepackage{enumitem}

\usepackage{etoolbox}

\newcommand{\spar}[1]{\vspace{1mm}{\noindent\bf #1.\hspace{2mm}}}

\begin{document}
\title{No Silk Road for Online Gamers!: Using Social Network Analysis to Unveil Black Markets in Online Games}
\author{Eunjo Lee, Jiyoung Woo, Hyoungshick Kim, Huy Kang Kim}
\thanks{E. Lee works at NCSOFT (e-mail: gimmesilver@ncsoft.com).}
\thanks{J. Woo works at Soonchunhyang University (e-mail: jywoo@sch.ac.kr).}
\thanks{H. Kim works at Sungkyunkwan University (e-mail: hyoung@skky.edu).}
\thanks{H.K. Kim works at Korea University (e-mail: cenda@korea.ac.kr).}

\begin{abstract}
Online game involves a very large number of users who are interconnected and interact with each other via the Internet.  
We studied the characteristics of exchanging virtual goods with real money through processes called ``real money trading (RMT)''. This exchange might influence online game user behaviors and cause damage to the reputation of game companies.  
We examined in-game transactions to reveal RMT by constructing a social graph of virtual goods exchanges in an online game and identifying network communities of users.

We analyzed approximately 6,000,000 transactions in a popular online game and inferred RMT transactions by comparing the RMT transactions crawled from an out-game market.
Our findings are summarized as follows: (1) the size of the RMT market could be approximately estimated; (2) professional RMT providers typically form a specific network structure (either star-shape or chain) in the trading network, which can be used as a clue for tracing RMT transactions; and (3) the observed RMT market has evolved over time into a monopolized market with a small number of large-sized virtual goods providers. 
\end{abstract}

%
%


\maketitle

\section{Introduction}
\label{section:introduction}
Online game is one of the most successful applications having a large number of users interacting in an online persistent virtual world through the Internet~\cite{infiniti:onlinegame}. Particularly, a massively multi-online role playing game (MMORPG) is considered one of the most popular online game genres. For example, there were about 11.7 million subscribers playing World of Warcraft, one of the most popular MMORPGs~\cite{lee2010server}. Users often compete against and/or collaborate with other users to complete given missions (called \emph{quest}) while playing an MMORPG. Online game companies have typically created various virtual goods (referred to as ``virtual currency'' or ``game items'') that are needed to enhance game character abilities. In most MMORPGs, virtual goods can legitimately be handed over from a user to another user as a trade or gift within the game. Similar to the real world, an economic system can be established in the virtual world so that virtual goods can be purchased, sold or exchanged between users~\cite{castronova2001virtual}. Moreover, users who do not want spend much time in playing a game but are eager to level up their characters might spend their real money in buying virtual goods. Such demands have led to the emergence of the exchanging market of virtual goods with real money called ``real money trading'' (RMT)~\cite{huhh2008simple}.
We note that RMT is different from ``In-app purchase'', which refers to the activities to purchase virtual goods in an official store.

Although RMT itself is a reasonable economic activity, the overproduction and popularity of RMT could have a negative impact on the virtual economy and cause severe damage to the reputation of the MMOPRG. In practice, professional cyber criminals known as a gold farming group (GFG)---a group of users running multiple client programs for MMORPGs on numerous machines to efficiently collect virtual goods---started to get involved in RMT transactions in a well-organized manner. Naturally, such activities by a GFG could discourage normal users from playing MMORPGs because their efforts are not properly rewarded~\cite{fujita2011detecting} and the game company experiences monetary damage as a result~\cite{castronova07:wow}. Furthermore, RMT is often associated with other criminal activities, such as money laundering~\cite{engage:group_laundering}, identity theft~\cite{woo2012automatic} and cheating~\cite{chen2005online}. Because of the double-sidedness of RMT, many online game companies have defined a policy to officially prevent RMT while some companies have launched an official RMT service~\cite{orland2017wow}. 

Besides, it is also important issue to monitor volumes and trends of RMT. RMT presents the potential revenue which online game company may get from In-app purchase. Consequently, the analysis of RMT could be used as important data for identifying potential consumers of virtual goods in the official market, understanding user preferences for virtual goods and managing the price of the virtual goods~\cite{lehdonvirta2008real}.

However, it is challenging to correctly identify RMT transactions from
a very large dataset in an MMORPG. When a user hands over a virtual item to another user for RMT, it is difficult to distinguish such an activity from legitimate activities (e.g., sending a virtual item to a friend as a gift). Further, players use the real money market outside the virtual world, so it is difficult to exactly match user identities in the real money market with accounts in the virtual world. 

To address those challenging issues, we focused on analyzing communities of users in a virtual goods trading network instead of individual RMT transactions between users. Game users, who are consumers in RMT, usually trade virtual goods with their friends and acquaintances rather than strangers, and as such, form a highly clustered structure in a virtual goods trading network. Moreover, previous studies~\cite{keegan2010dark, woo2011can} demonstrated that members in a GFG, who are main providers in RMT, are also likely to form a separate community in a virtual goods trading network because a GFG member is obligated to frequently perform trading activities with another member in the same GFG. That is, the trading activities between different communities, especially provider communities and consumer communities, may represent RMT transactions, while the trading activities between users within a community may represent legitimate trading activities.

Recent advances in the field of social network analysis provided us with the fundamental tools to detect communities in a network and understand their characteristics better. We used approximately 6,000,000 virtual goods trading transactions between April 2015 and June 2016, collected from more than 200,000 users in Lineage\footnote{\url{http://lineage.plaync.com/}}, one of the most popular MMORPGs in Korea, to construct a virtual goods trading network. We analyzed its key characteristics utilizing nodes to represent users and edges to represent trading transactions (e.g., sending an item) between users, respectively. Our key contributions are summarized as follows:

First, we detected RMT groups and estimated the size of RMT market (e.g., the number of RMT transactions) in an online game using community detection algorithms in a virtual goods trading network and evaluated our estimation with the dataset crawled from two real world RMT websites. This is a significant advancement from previous literature~\cite{fujita2011detecting} because we divided a given communities into more fine-grained categories based on network structure, and analyzed users' play styles and in-game economic activities to identify the target groups accurately while the previous study used trade activities only.

Seconds, we analyzed the key characteristics of the network structure and found that star-shaped and chained structures were typically formed for professional RMT provider groups.

Third, we analyzed the dynamics of virtual goods trading behaviors over time. We constructed weekly trading networks from virtual goods trading transactions during a year. We then demonstrated how a virtual goods trading network grows and changes over time by visualizing the temporal changes of the network. Finally, we observed that the market has evolved into a monopolized market with a small number of RMT provider groups.

\section{RMT Flow}
\label{section:Background}
RMT is an interesting economic activity because it is used to exchange assets in a virtual world with real world value~\cite{huhh2008simple}. As a result of increasing demands from users for trading virtual goods, RMT markets have naturally become popular in recent years, although RMT is prohibited in most online game companies~\cite{heeks2010real}. In addition, by the nature of the black market, professional criminals (i.e., GFG) have, in turn, materialized in the RMT market~\cite{heeks2008current}. The size of the RMT market is getting larger and larger, and now, is becoming a new industrial market. 
To summarize, the overproduction of RMT is one of the serious problems that can cause imbalance of game economy and inequality of wealth and opportunity. It exploits the economic safety system and destroy the overall design of player's growth model.

\begin{figure}[ht]
\centering
\includegraphics[width=8.5cm]{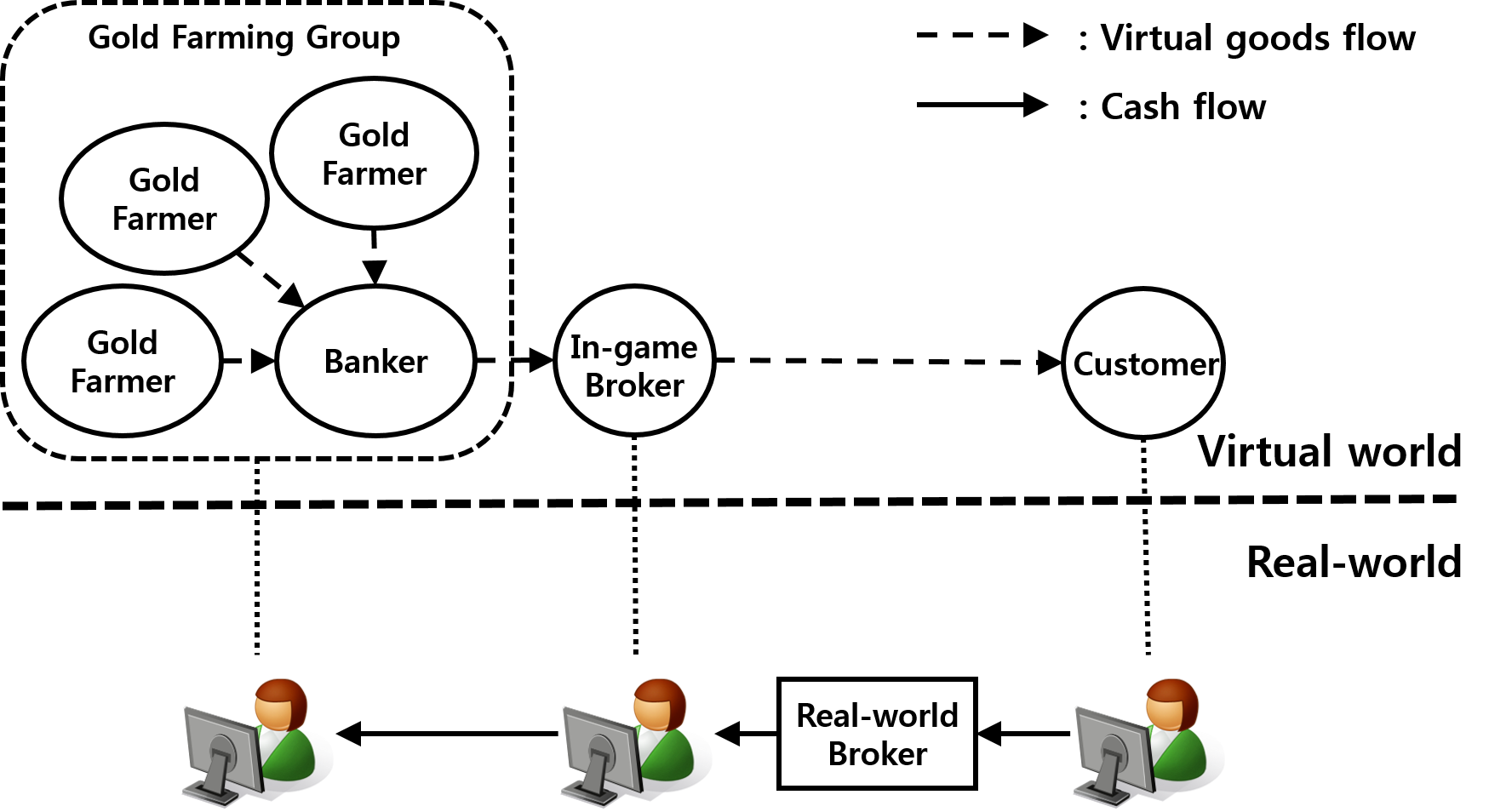}
\caption{Typical flow of cash and virtual goods for RMT.}
\label{fig:rmt flow}
\end{figure}

Fig.~\ref{fig:rmt flow} briefly describes how GFG is organized and connected to consumers for RMT. A typical GFG consists of a few \emph{bankers} and many \emph{gold farmers} in a hierarchical manner~\cite{kwon2013surgical}. A gold farmer is an entity working full time playing online games in order to harvest virtual goods by using automated programs (e.g., game bots and macros) or by hiring low-cost laborers~\cite{lee2016you}. Several gold farmers work together as a team to accomplish challenging tasks such as hunting a gigantic monster. A banker is an entity who gathers virtual goods from the gold farmers in his or her own GFG and sells it to online customers.

Brokers have emerged with the growing popularity of commercial MMORPGs (e.g., ``Ultima Online'', ``World of Warcraft'' and ``Lineage''). There are two types of brokers in the RMT transactions. One is an in-game broker who performs intermediate trades between GFGs and customers inside the virtual world. 
Another is a real world broker (i.e., RMT website) who advertises massive quantities of virtual goods to customers and provides a reliable payment system (e.g., escrow). In most cases, brokers receive a commission from the sellers or buyers in RMT. 

\section{Dataset}
We use a dataset of Lineage, which is released by NCSOFT, in this study. It is a popular MMORPG in Korea with a long history, and it has also English, Chinese and Japaneses versions. Various activities performed by the user in the virtual world or state change events are recorded in the form of a structured log through the game server in Lineage. The game log consists of the following information.

\begin{itemize}[leftmargin=.5cm]
\item Time: When a specific activity or event occurs
\item Location: Where a specific activity or event occurs
\item User information: the user's character ID, level, job and race, etc.
\item Event ID: An ID that identifies the type of activity (hunting, trading, item acquisition, dungeon entry, etc.) or event type (level up, game money increasing or decreasing, etc.)
\item Target user information: The other character information if the log is related to the interaction with another user
\item Detailed information: Depending on each activity type or event type, detailed information related to the activity type or event type
\end{itemize}

Specifically, we used the game activity logs of more than 200,000 users from April 2015 to June 2016. 

\section{Trade Network Analysis for RMT Detection}
\label{section:experiments}
We propose a framework highlighting five phases to detect RMT: (1) Given a set of virtual goods transactions in game log files, we construct a virtual goods trading network, and then find community structures in the constructed network; (2) Using the characteristics of the structure of the community, we categorize the communities into five groups; (3) We categorize game users into seven groups by play style in virtual world; (4) After analyzing the network characteristics of each community type and behavior types of users belonging to the community, we diagnose abnormal communities and detect suspected RMT group. The community groups are tagged into provider type, consumer type and non-RMT type; (5) When we classify communities into provider, consumer and others groups, transactions between provider and consumer communities are filtered out as RMT.

\subsection{Construction of a virtual goods trading network and community detection}
\label{section:trade construction}

In general, there are two types of trades in online games: (1) \emph{direct} trade and (2) \emph{indirect} trade, by which one deposits goods in the warehouse and then another retrieves it.

Given a set of trade transactions in an online game, we construct a \emph{weighted} directed network called ``virtual goods trading network''. In this network, a node represents a character (i.e., account) or a warehouse and an edge represents a virtual goods trade. The weight of an edge reflects the number of trades between them. We constructed a virtual goods trading network on a weekly basis because Lineage has a weekly shutdown for its content updates and server maintenance.

Lineage operates several game servers. Although a user has several characters, one character should play on one game server. All servers are disconnected, and all characters can trade items with just others in the same server.
Among fifty operating game servers in Lineage, we chose one server in which RMT appears to be the most active. 
From the game logs in the server, we constructed a virtual goods trading network. Table~\ref{table:trading network summary} summarizes some basic properties of the constructed virtual goods trading network. 

\begin{table}[ht]
\centering
\caption{Summary of network characteristics of virtual goods trading network.}
\label{table:trading network summary}
\begin{tabular} {|c|c|c|c|} \hline
Feature & Value & Feature & Value \\\hline
\# of nodes & 25,633 & degree mean & 2.125 \\\hline
\# of edges & 36,598 & degree std. & 10.8338 \\\hline
average path length & 8.8283 & betweenness mean & 31,709 \\\hline
clustering coefficient & 0.001989 & betweenness std. & 464,528 \\\hline
\end{tabular}
\end{table}

When a virtual goods trading network is established, we try to find \emph{communities} in the trading network where nodes in a community construct denser relationships than between nodes in the remaining communities. This is a reasonable technique to detect suspicious trading behaviors---people usually trade virtual goods with their friends and acquaintances rather than strangers, and thus form a highly clustered structure in a virtual goods trading network. 

To achieve this here, we tested five community detection algorithms (`multilevel~\cite{blondel2008fast}', `fastgreedy~\cite{clauset2004finding}', `walk trap~\cite{pons2005computing}', `label propagation~\cite{raghavan2007near}' and `infomap~\cite{rosvall2007maps}') and compared their performances based on the \emph{modularity} that is a quantity that represents how well communities are constructed~\cite{clauset2004finding}. As more densely connected communities are formed, the modularity closes to one. 

From the results of modularity tests, we selected the `multilevel' algorithm. This algorithm has the highest modularity value (see Table~\ref{table:community detection performance}).

\begin{table}
\centering
\caption{Comparison of community detection algorithms in terms of modularity.}
\label{table:community detection performance}
\begin{tabular} {|c|c|c|c|} \hline
\multirow{2}{*}{Algorithm} & \multicolumn{3}{c|}{Modularity} \\\cline{2-4}
 & Min. & Avg. & Max. \\\hline
multilevel & \textbf{0.73} & \textbf{0.97} & \textbf{0.99}  \\\hline
fast greedy & 0.71 & 0.96 & 0.99 \\\hline 
walk trap & 0.61 & 0.75 & 0.87  \\\hline
label propagation & 0.61 & 0.91 & 0.96 \\\hline
infomap & 0.66 & 0.91 & 0.94 \\\hline
\end{tabular}
\end{table}

\begin{figure}[ht]
\centering
\begin{tabular}{c c} \\
\includegraphics[width=3.5cm]{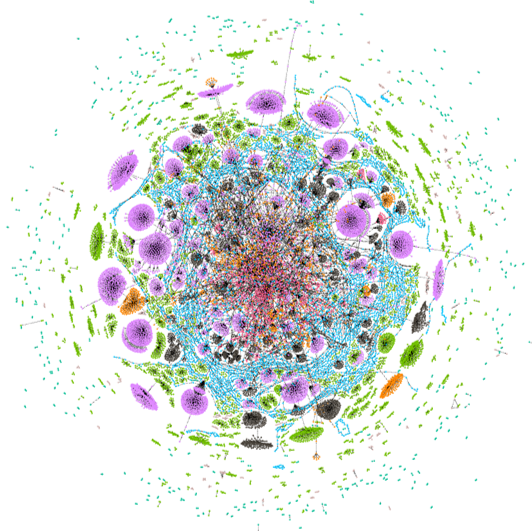} & 
\includegraphics[width=3.5cm]{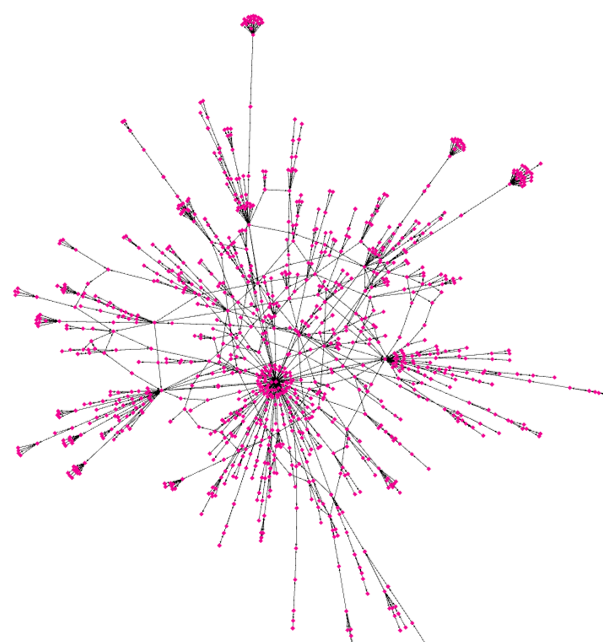} \\
(a) Total network & (b) Type 1 \\
\includegraphics[width=3.5cm]{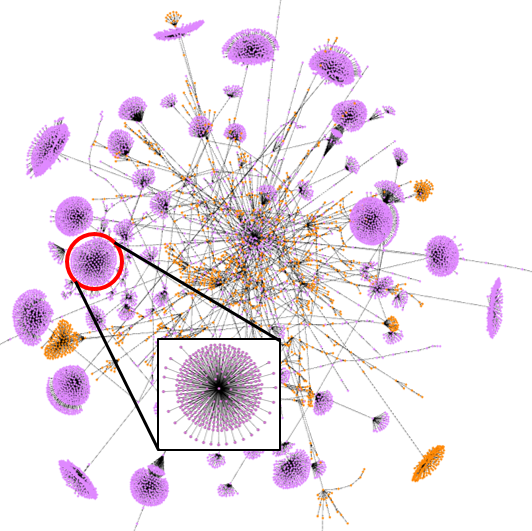} & 
\includegraphics[width=3.5cm]{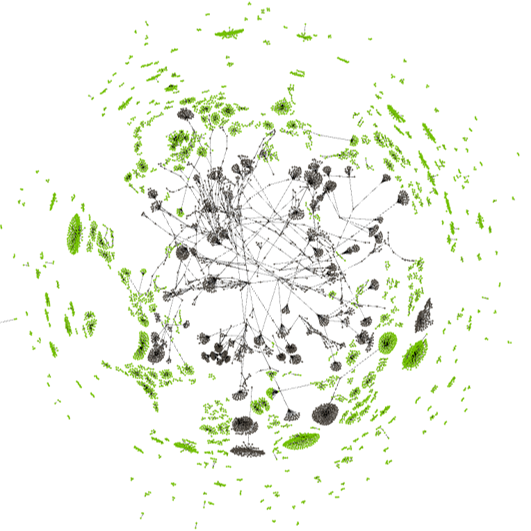} \\
(a) Type 2 & (b) Type 3 \\
\includegraphics[width=3.5cm]{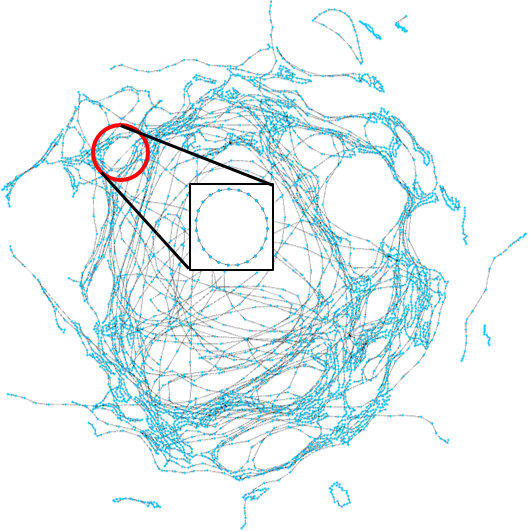} &
\includegraphics[width=3.5cm]{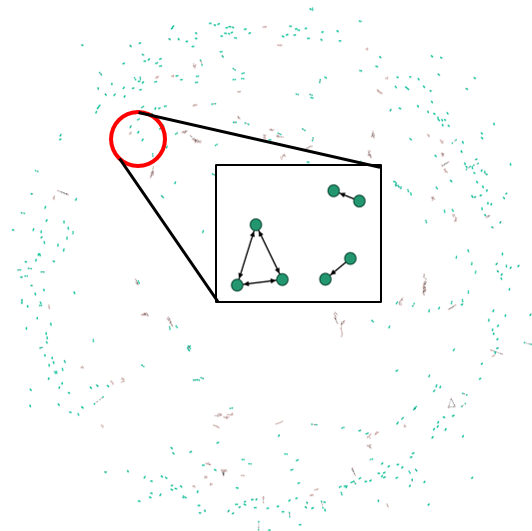}\\
(c) Type 4 & (d) Type 5 \\
\end{tabular}
\caption{Visualization of community types}
\label{fig:clustered network}
\end{figure}

\begin{figure}[ht]
\centering
\begin{tabular}{c c} \\
\includegraphics[width=3.8cm, height=2.5cm]{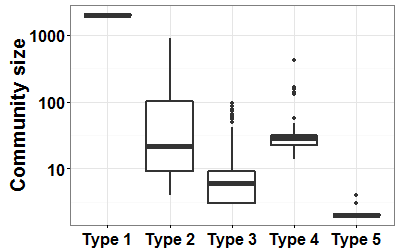} & \includegraphics[width=3.8cm, height=2.5cm]{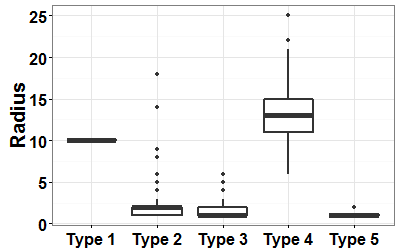}\\
(a) Community size & (b) Radius \\
\includegraphics[width=3.8cm, height=2.5cm]{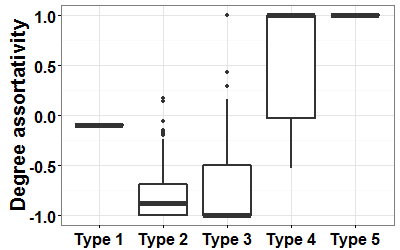} & \includegraphics[width=3.8cm, height=2.5cm]{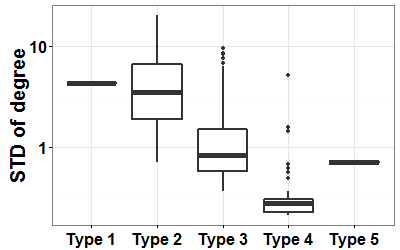}\\
(c) Degree assortativity & (d) STD of degree\\
\end{tabular}
\caption{Network characteristics of community types.}
\label{fig:community statistics}
\end{figure}

\subsection{Community grouping by network structure}
\label{section:categorization}
 
When community structures are found in a trading network, we categorize those communities into several interesting groups by their network statistics using the $k$-means clustering algorithm. We used the following statistics:

\spar{Mean and standard deviation of degree centrality} The degree is the number of connections that a node has to other nodes. The degree of the community characterizes the network in such a way that the average degree is closely related with the network density and the standard deviation shows how evenly the degree centrality is distributed, in other words, centralization of the network. 

\spar{Mean and standard deviation of betweenness} We adopt the betweenness centrality to show the centralization in terms of information flow. The betweenness centrality of the node implies the ability to broker between groups. 
The betweenness centrality of a node is calculated as the number of shortest paths from all nodes to others through the target node. These statistical features imply that many members have the capability to efficiently deliver the asset or information within the community. 

\spar{Degree assortativity coefficient} 
Degree assortativity is a preference of the node to attach to others that are similar in terms of degree. 
The value will be close to one if two or more connected nodes have the same degree, or negative one otherwise.

\spar{Clustering coefficient} A clustering coefficient is a measure of the likelihood that three nodes will become fully connected and form a clique. We adopt the clustering coefficient as a measure of the property that all members in the community have a trade-relationship with each other. 

\spar{Radius} The radius is the longest distance path from all paths between nodes in the community. This metric shows how long it takes to deliver information or assets to all members in a community.

\spar{Community size} To characterize the community in their scale, we use the number of nodes in a community to derive the community size.

\vspace{0.2cm}

According to our experiments, trading communities are clustered into five groups. Table~\ref{table:community count} shows the number of communities and the ratio of community members in the total number of users participating in the virtual goods trading by community type (type 1, 2, 3, 4 and 5). Fig.~\ref{fig:clustered network} shows a visualization of the five categorized communities. We derive unique network topology by clustering communities in terms of network statistics as mentioned above. Fig.~\ref{fig:community statistics} shows the differences in the network statistics by community types.

\begin{table}[ht]
\caption{Community count and community member's rate for community types}
\centering
\begin{tabular}{|c|c|c|c|c|c|} \hline
Type & 1 & 2 & 3 & 4 & 5 \\\hline
\# of community & 1 & 19 & 732 & 104 & 471 \\\hline
Member rate & 10.21\% & 37.00\% & 33.38\% & 15.62\% & 3.78\% \\\hline
\end{tabular}
\label{table:community count}
\end{table}

\subsection{User grouping by play style}
To further characterize communities, we investigate users' play style per communities.
We summarized users' play data using features which are described in table~\ref{table:user clustering features} and categorized users into seven groups by play style using $k$-means clustering algorithm (see Table~\ref{table:play pattern}).
Table~\ref{table:proportion of play style} shows the proportions of user play styles in each community type.

\begin{table}[ht]
\caption{Features for user clustering.}
\centering
\scalebox{0.9}{\renewcommand{\arraystretch}{1.2}%
\begin{tabular}{|c|c|c|} \hline
No. & Feature \\\hline
1 & \parbox{8cm}{Play time (unit: second)} \\\hline
2 & \parbox{8cm}{The number of days a user plays} \\\hline
3 & \parbox{8cm}{The amount of experience a user obtains}  \\\hline
4 & \parbox{8cm}{The number of death}  \\\hline
5 & \parbox{8cm}{The number of combats against another player} \\\hline
6 & \parbox{8cm}{The number of combats against monster} \\\hline
7 & \parbox{8cm}{The number of dungeons a user plays} \\\hline
8 & \parbox{8cm}{The number of parties a user joins} \\\hline
9 & \parbox{8cm}{The number of an item was enchanted by a user} \\\hline
10 & \parbox{8cm}{The number of an item was traded with another user} \\\hline
11 & \parbox{8cm}{The amount of game money a user gets from other users} \\\hline
12 & \parbox{8cm}{The amount of game money a user gives to other users} \\\hline
13 & \parbox{8cm}{The amount of money a user obtains} \\\hline
14 & \parbox{8cm}{The amount of money a user spends} \\\hline
15 & \parbox{8cm}{The amount of time a user spends for fishing} \\\hline
16 & \parbox{8cm}{The amount of time a user spends for shopping} \\\hline
\end{tabular}
}
\label{table:user clustering features}
\end{table}

\begin{table}[ht]
\centering
\caption{Categorization of users by play style.}
\label{table:play pattern}
\scalebox{0.85}{
\begin{tabular} {|c|c|} \hline
Play style & Description \\\hline
Fisher & users who focus to play for fishing contents \\\hline
Genuine & users who enjoy various contents  \\\hline
\multirow{2}{*}{Hard core} & users who have very high play time and \\\ & various activities extremely \\\hline
Light & users who have low play time and activities\\\hline
Party & users who focus to party-activities \\\hline
Shop host & users who play shop activities heavily \\\hline
\multirow{2}{*}{Worker} & users who concentrate on activities for \\\ & earning game money, i.e., hunting, harvesting \\\hline
\end{tabular}
}
\end{table}

\begin{table}[ht]
\caption{Proportions of users' play styles in each community type.}
\label{table:proportion of play style}
\scalebox{0.85}{
\begin{tabular} {|c|r|r|r|r|r|} \hline
Play style & Type 1 & Type 2 & Type 3 & Type 4 & Type 5 \\\hline
Fisher & 13.12\% & 0.01\% & 3.83\% & 0.14\% & 16.73\% \\\hline
Genuine & \textbf{46.67\%} & 0.07\% & 18.42\% & 0.54\% & \textbf{31.99\%} \\\hline
Hard core & 0.55\% & 0.14\% & 0.22\% & 0.32\% & 0.40\% \\\hline
Light & 21.69\% & \textbf{75.25\%} & 25.79\% & 0.30\% & 26.37\% \\\hline
Party & 2.43\% & 0.00\% & 5.97\% & 0.16\% & 2.81\% \\\hline
Shop host & 14.21\% & \textbf{24.33\%} & \textbf{21.87\%} & \textbf{98.54\%} & 21.15\% \\\hline
Worker & 1.32\% & 0.20\% & \textbf{23.89\%} & 0.00\% & 0.54\% \\\hline
\end{tabular}
}
\end{table}

\subsection{RMT group detection}
\label{section:community tagging}
After clustering communities and users, we compare users' in-game economic activities and the proportion of types of users in each community type. 

\begin{figure}[ht]
\centering
\begin{tabular}{c c} \\
\includegraphics[width=3.8cm]{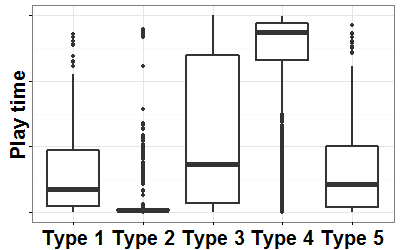} &
\includegraphics[width=3.8cm]{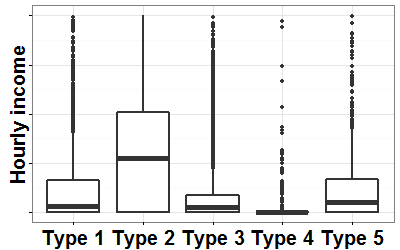} \\
(a) Play time & (b) Hourly income \\
\includegraphics[width=3.8cm]{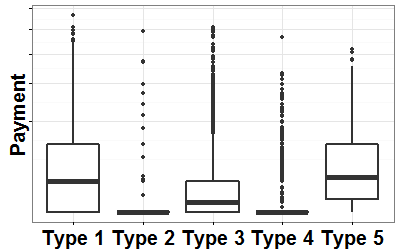} &
\includegraphics[width=3.8cm]{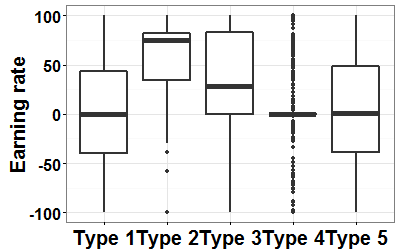} \\
(c) Payment to company & (d) Earning rate/consuming \\
\includegraphics[width=3.8cm]{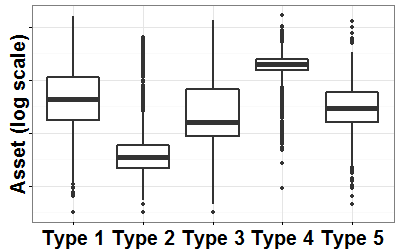} &
\includegraphics[width=3.8cm]{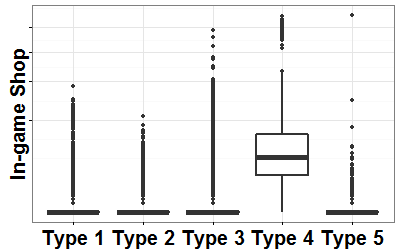} \\
(e) In-game asset & (f) In-game shop activity
\end{tabular}
\caption{Users' in-game economic activities by community types.}
\label{fig:boxplot of cluster}
\end{figure}

According to the previous study~\cite{heeks2009understanding}, providers have a unique play style to achieve efficiency. Thus, the analysis on play styles could be used to detect RMT providers and consumers (see Table~\ref{table:proportion of play style}).
We also analyze users' in-game economic activities in each community type (see Fig.~\ref{fig:boxplot of cluster}). In most cases, the purpose of RMT providers is only to make money so they did not pay to game companies, while consumers often pay to game company for their fun. Consequently, the differences in the amount of economic activities could be used to categorize community types into consumer group and provider group.
The users in normal communities buy items from RMT providers, thus we call the normal community the consumer community.
Based on the results in Table~\ref{table:proportion of play style} and Fig.~\ref{fig:boxplot of cluster}, we will explain the main characteristics of each community type and discriminate communities into types in the following sections.

\subsubsection{Type 1: giant component}
\label{section:giant component}
The members in this type comprise the largest community (see Fig.~\ref{fig:community statistics}(a)) and have the most inter-community trades (see the center node in Fig.~\ref{fig:inter community network}). Particularly, the giant component has characteristics of a scale-free network by which degree distribution follows a power-law. In a scale-free network, the probability distribution of degree complies with Eq.~\ref{eq:power law}. 
\begin{equation}
\label{eq:power law}
p(x) = Cx^{-\alpha}
\end{equation}

According to this test for the community of type 1, $\alpha$ for giant component ranged from 2.7 to 2.8.

The giant component also has the most `genuine' players who enjoy various game contents (see Table~\ref{table:proportion of play style}), and pay the most to the game company among communities (see Fig.~\ref{fig:boxplot of cluster}(c)). 
From above observations, we conclude that this community is naturally formed from normal users' trades and is a consumer group in RMT.

\subsubsection{Type 2: large star-shaped network}
\label{section:large star}
This type is a community in which a few nodes are in the center and remaining nodes trade only with center nodes. The degree of most nodes is one or two, while the degree of the center nodes is in the hundreds. Consequently, the standard deviation of the degree in this type is very high, while degree assortativity is close to -1 (see Fig.~\ref{fig:community statistics}(c) and (d)). Previous studies~\cite{kwon2013surgical, woo2011can,kwon2017crime} showed that gold farming groups tend to construct communities similar to this type. The reasoning why this type is a gold farming group follows. Gold farmers repeatedly earn game money and items. Collected items and game money are delivered to merchant characters, and the merchant characters sell the items for game money. The game money from gold farmers and the money acquired through item trade by a merchant character flow to banking characters~\cite{kwon2017crime}.
From this point of view, the center node is a banker whose role is to gather and store virtual assets for GFGs.

Characters in these communities focus on producing virtual assets, however, do not consume the virtual asset in their play as shown in Fig.~\ref{fig:boxplot of cluster}(d)\footnote{$(p-c)/(p+c),~p:production,c:consumption$}. However, they do not possess many assets because they frequently send bankers or consumers their production (see Fig.~\ref{fig:boxplot of cluster}(e)). Moreover, their payment for the game company is lower than other communities because their purpose of game playing is economic income by their labor, not entertainment (see Fig.~\ref{fig:boxplot of cluster}(c)).

These communities are mainly composed of `Shop host' and `Light' type users. `Shop host' users are merchant characters and `Light' users are gold farmers, who have a short play time to avoid game bot detection. They conforms peripheral nodes from the perspective of the network structure in Fig.~\ref{fig:clustered network} (c). 

\subsubsection{Type 3: small star-shaped network}
The network structure of type 3 is similar to type 2, but its size is smaller than type 2. The interesting point is that the play style of this type is different from type 2 in terms of play time and productivity even though the network structure is similar to type 2. This type consists of players with more various play styles than type 2 (see Table~\ref{table:proportion of play style}). 
In the economic point of view, type 3 seems to be less efficient than type 2. Type 2 pays less for the game company, however, produces more income than type 3. That is, users in the large star earn more game money hourly with less play time and cost (see Fig.~\ref{fig:boxplot of cluster}(a), (b) and (c)). Besides, the total revenue of type 2 is much higher than type 3. 
Consequently, we guess that communities of type 3 are unspecialized and immature RMT providers, while communities of type 2 are professional organizations.

\subsubsection{Type 4: chain network}
\label{section:chain network}
The criminal network is organized to balance efficiency and security~\cite{morselli2007efficiency}. The centralized network has a high efficiency, but weak for targeted attack ~\cite{albert2000error}. According to previous works,  decentralization has also been observed to be a key tactic adopted by members of a criminal network to reduce the risk of targeting and asset confiscation ~\cite{morselli2007law}. We could also observe this decentralization in MMORPGs. 

This type is a very different structure from the star network and looks like peer-to-peer network. Most nodes in this community have one or two degrees, and their connections construct a long circular chain. 
A node in this type sends all items to another node and then, after a time the receiving node sends all of these assets back to the other. These trading activities form a circular chain. Consequently, these communities have a high radius and degree assortativity (see Fig.~\ref{fig:community statistics}(b) and (c)).
Regarding economic activities, the characters' focus is on playing for a long time (see Fig.~\ref{fig:boxplot of cluster}(a)), however, they exclude producing or consuming activities (see Fig.~\ref{fig:boxplot of cluster}(b) and (d)). Moreover, they pay minimal amounts to the game company (see Fig.~\ref{fig:boxplot of cluster}(c)). They mostly perform an in-game shop activity primarily buying or selling items to other characters in the virtual world (see Fig.~\ref{fig:boxplot of cluster}(f)). Based on these observations, we could conclude this group is an abnormal group.

We discovered that these are arbitrage groups, who realize profit from buying virtual items and selling them back. They possess a large amount of assets in order to buy bulky items at a low price, thus this type of community accounts for the largest asset portions in the game world (see Fig.~\ref{fig:boxplot of cluster}(e)). This unique type community also appears because of the game rule.

In MMORPGs, the cost to buy and sell items, i.e., real estate rental fee, wages, distribution costs, etc, is minimal. Consequently, a player can easily make money from arbitrage if enough in-game currency and time is available. However, there is one limit for arbitrage in Lineage. Lineage imposes a duty on characters if they participate in shop activity more than forty times in a week to prevent massive in-game shop activities. Eventually, a character in an arbitrage group does hit the shop activity limit and sends another character all the virtual goods obtained. The character receiving the goods continues with the arbitrage. After a week, the original character receives all items and returns to arbitrage.
Consequently, the chain network is constructed to reduce the risk of target banning by the game company and avoid the constraint on economic activities. We could conclude that these communities are RMT providers as well, because they play to get a benefit like GFG.

\subsubsection{Type 5: outcast communities}
The play style of this type is similar to that in type 1 (see Fig.~\ref{fig:boxplot of cluster} and Table.~\ref{table:proportion of play style}). However, users in this community have trades with only one or two friends. We conclude that users in this type are normal users who limit their social relationship regarding trading. We filtered out their trades from our estimation.

\subsubsection{Evaluation of RMT group detection}

Based on the analysis of this step, we could detect abnormal community types and name abnormal communities as provider of RMT by network structure. 
To evaluate the detecting result, we measured the rate of users who were banned by using game bot among users belonging to each community type. Table~\ref{table:ban user} shows that RMT provider groups have more banning users than other groups. 

\begin{table}[ht]
\centering
\caption{The rate of banning users by community types}
\label{table:ban user}
\begin{tabular} {|c|c|c|c|c|} \hline
Type 1 & Type 2 & Type 3 & Type 4 & Type 5 \\\hline
10.05\% & \textbf{65.60\%} & \textbf{23.01\%} & \textbf{10.64\%} & 7.64\% \\\hline
\end{tabular}
\end{table}

\subsection{RMT Estimation}
\label{section:rmt estimation}

\begin{figure}[ht]
\centering
\includegraphics[width=6cm]{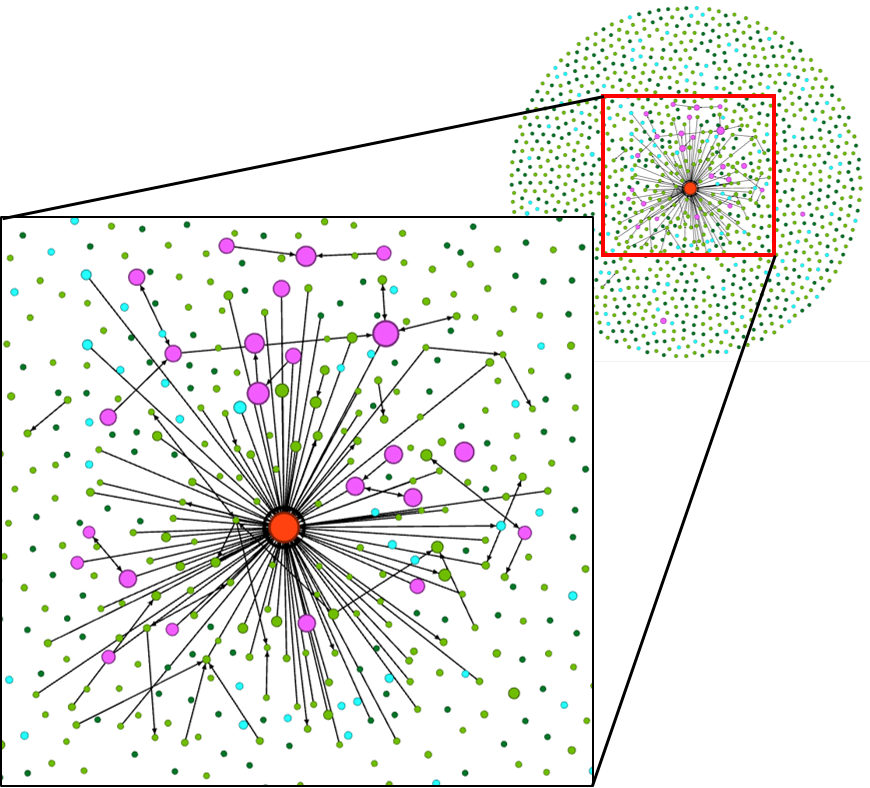} \\
\caption{Inter-community trading network with nodes (communities) and edges (inter-community trade transactions).}
\label{fig:inter community network}
\end{figure}

In this step, we built the inter-community trading network where nodes and edges represent communities and inter-community trade transactions, respectively (see Fig.~\ref{fig:inter community network}). After that, we extracted trading transactions between provider communities and consumer communities as RMT, which reaches 5\% of the total transactions.

To compare with the RMT estimation, we crawled publicly accessible RMT records between May 2015 and June 2016 from two popular RMT websites (ItemBay\footnote{\url{http://www.itembay.com}} and ItemMania\footnote{\url{http://www.itemmania.com}}). 
According to the web traffic ranking website named Rankey\footnote{\url{https://www.rankey.com/}}, ItemBay and ItemMania dominate the Korean RMT market, with approximately 90\% of the total traffic volume for RMT websites in Korea. A typical RMT record is composed of unit price, trade volume, game server and trade completion time.

The best way to evaluate the performance of our estimation is to measure the precision and recall rates using the real RMT records at RMT websites as the ground truth. When real transaction records are not accessible, we can only approximate RMT transactions by articles posted in RMT sites, but the approximation naturally has error because of following reasons. 

First, the trading time in RMT websites and in the game world does not exactly coincide because the trades take place in the virtual world and real world at different times. To minimize the error from this, we used weekly based time window. If we use small time windows (e.g. daily based, or hourly based) then it can miss RMT record occurred in the very next time period. Setting long-term time window for the analysis can increase the possibility of including RMT events occurred in a short time difference.

Second, the actual transaction volume is not precisely equal to the volume posted on the RMT website. The volume in the article is the maximum amount that the seller wants to trade, however, most customers buy only part of it. We do not have the exact information available on the transaction volume in the RMT website.

Third, there is no confirmation that the user in RMT websites is the same user in the game even if they used the same user name.

As an alternative to the problem, we computed the correlation coefficient of weekly trade transactions between the RMT transactions estimated by the proposed method and those of real RMT records at RMT websites.

For our analysis, we categorized trades into three categories as follows:

\spar{Intra} The trades between users in an identical community.

\spar{Inter} The trades between users in different communities.

\spar{Inter (RMT)} The trades between RMT providers and consumers.

\vspace{0.2cm}

As we might expect, the trade transactions of ``Inter (RMT)'' and RMT websites have a statistically significant correlation, while the correlation coefficients for other categorized transactions do not (see Table~\ref{table:correlation of transactions}).

\begin{table}[ht]
\centering
\caption{Correlation coefficient $\rho$ of the number of transactions per week with RMT websites (*: 0.05, **: 0.01, ***: 0.001).}
\label{table:correlation of transactions}
\begin{tabular} {c|c|c|c|} \cline{2-3}
 & $\rho$ & p-value \\\hline
\multicolumn{1}{|c|}{Intra} & -0.1280  & 0.3754 \\\hline
\multicolumn{1}{|c|}{Inter} & 0.0082  & 0.9534 \\\hline
\multicolumn{1}{|c|}{Inter (RMT)} & 0.4722 & 0.000468*** \\\hline
\end{tabular}
\end{table}

\section{Measurement study of RMT}
\subsection{Volume of RMT}
As we mentioned in the previous section, the volume of transactions crawled from RMT websites could overestimate the real transactions. 

\begin{table}[ht]
\centering
\caption{Volume of virtual goods per transaction (Qu.: Quartile, unit: game money).}
\label{table:summary RMT currency}
\begin{tabular} {|c|c|c|c|} \hline
Category & Intra & Inter (RMT) & RMT website \\\hline
Min. & 1 & 50,000 & 500,000 \\\hline
1st Qu. & 25,290 & 5,000,000 & 11,000,000 \\\hline
Median & 1,500,000 & 12,000,000 & 30,000,000 \\\hline
3rd Qu. & 6,315,000 & 32,620,000 & 54,380,000 \\\hline
Max. & 1,276,000,000 & 300,000,000 & 860,000,000 \\\hline
Mean & 10,310,000 & 28,520,000 & 57,640,000 \\\hline
STD & 36,335,270 & 44,811,145 & 95,322,270 \\\hline
\end{tabular}
\end{table}

\begin{figure}[ht]
\centering
\includegraphics[width=8cm, height=4cm]{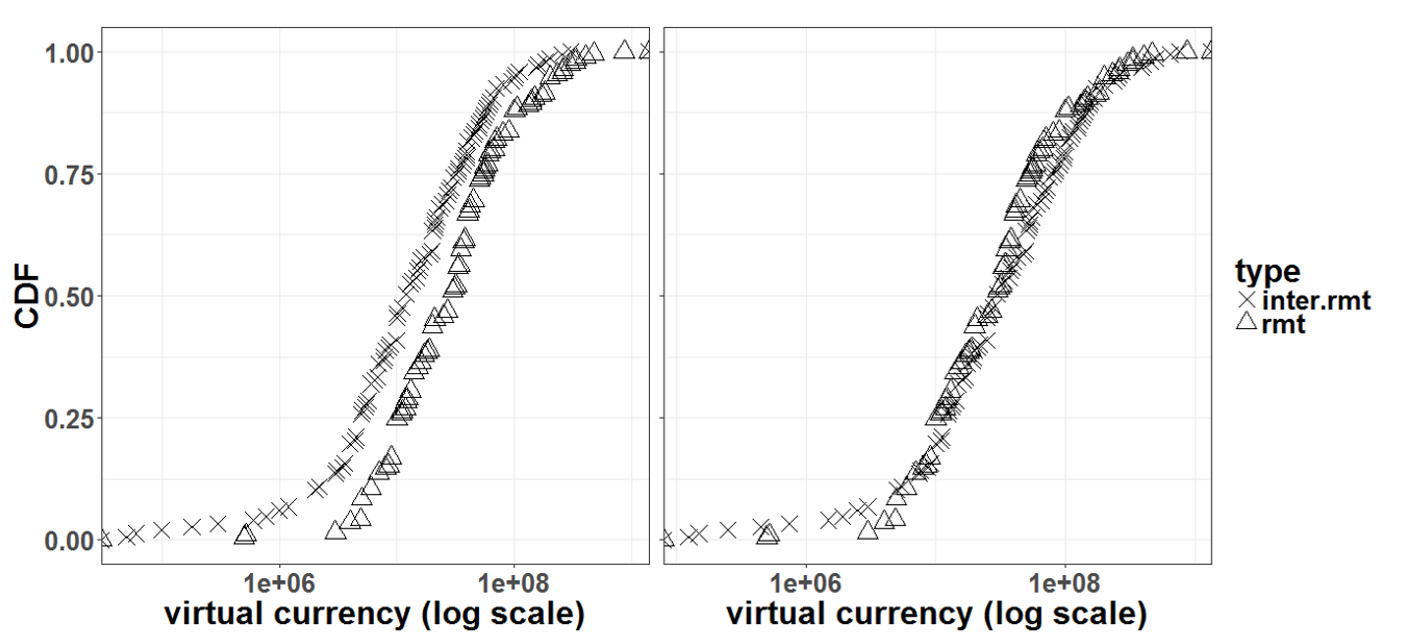}
\caption{CDF of the volume of virtual goods per a transaction of inter-RMT and RMT website(left: original data, right: normalized data).}
\label{fig:cdf}
\end{figure}

\begin{figure}[ht]
\centering
\includegraphics[width=7cm, height=4cm]{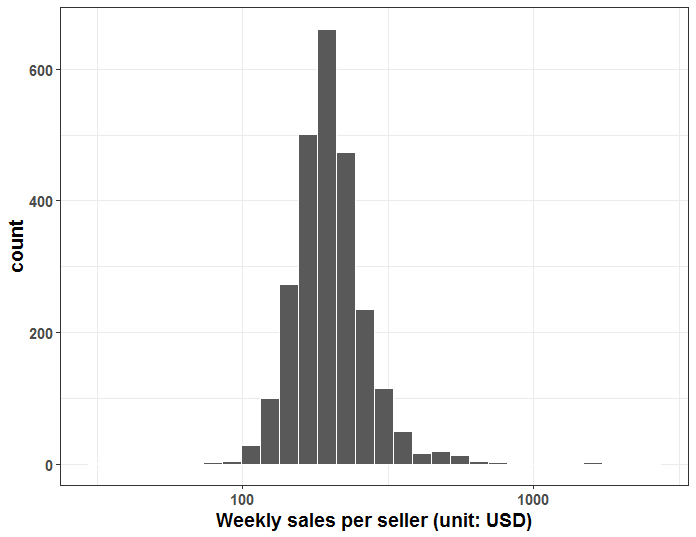}
\caption{histogram of the weekly sales per game account of RMT provider.}
\label{fig:rmt sales}
\end{figure}

To get information on RMT volume, we analyzed the distribution of individual transaction volume. The actual RMT transaction volume from our estimation reaches approximately 40\% of the transaction volume in RMT websites (see Table~\ref{table:summary RMT currency}). Fig.~\ref{fig:cdf} shows that the distribution of the transaction volume in the RMT websites is close to the distribution of our RMT estimation if we perform the \emph{median normalization} for the transaction volume in RMT websites.

The RMT website shows a cash price for the game money, so we can use it to estimate the volume of seller's sales. According to our estimations, the total volume of RMT for Lineage is approximately sixty millions USD per year. This is equivalent to about 25\% of Lineage's annual sales in 2014. The seller's average weekly sales is about 209 USD (see Fig.~\ref{fig:rmt sales}). Among the things that we have seen, the highest total volume per game account recorded sales of about half million USD per year. Given the fact that RMT providers usually operate multiple game accounts, the actual average sales per RMT provider are likely to be larger.

\subsection{Monopoly of RMT market}
\label{section:dynamics of RMT networks}
In this section, we analyze how the RMT market changes over time. Lineage consists of multiple servers; each server manages its own virtual world with different history. The oldest world has been in service for eighteen years while the youngest world has been in service for only half a year. Interestingly, we can see that the structures of trading networks in those servers are significantly different with elapsed time.

\begin{figure} [ht]
\centering
\begin{tabular} {c c} \\
\includegraphics[width=4cm]{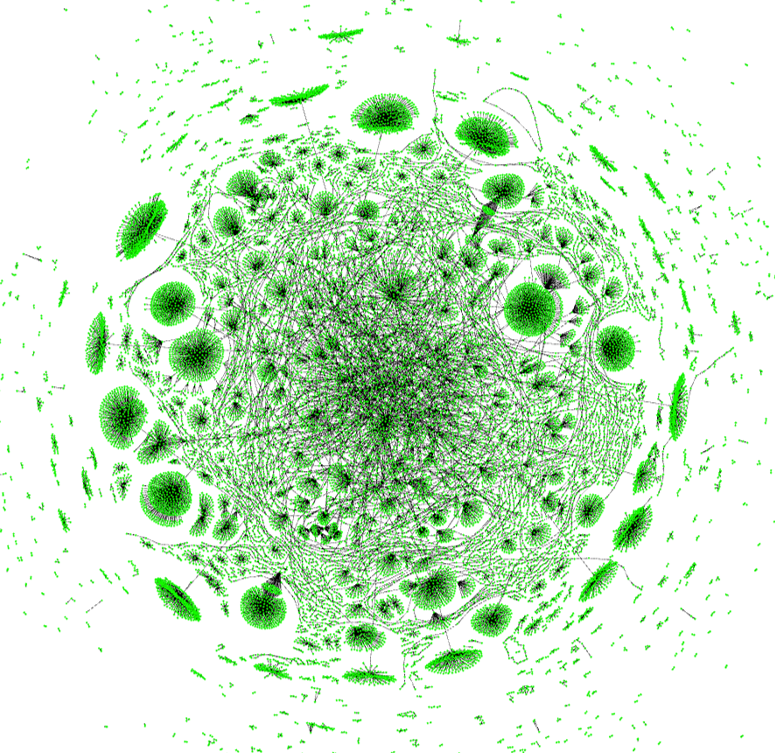} &
\includegraphics[width=3.5cm]{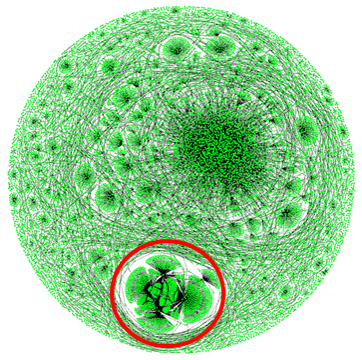} \\
(a) Young server & (b) Old server \\
\end{tabular}
\caption{Comparison of network structures between young server and old server.}
\label{fig:comparison old and young}
\end{figure}

In Fig.~\ref{fig:comparison old and young}, we can observe that many provider communities compete with each other for an extensive consumer community in the young server (see Fig.~\ref{fig:comparison old and young}(a)) while a few extra-large provider communities monopolize RMT in the old server (see the red circle in Fig.~\ref{fig:comparison old and young}(b)).

\begin{figure}[ht]
\centering
\includegraphics[width=7.5cm]{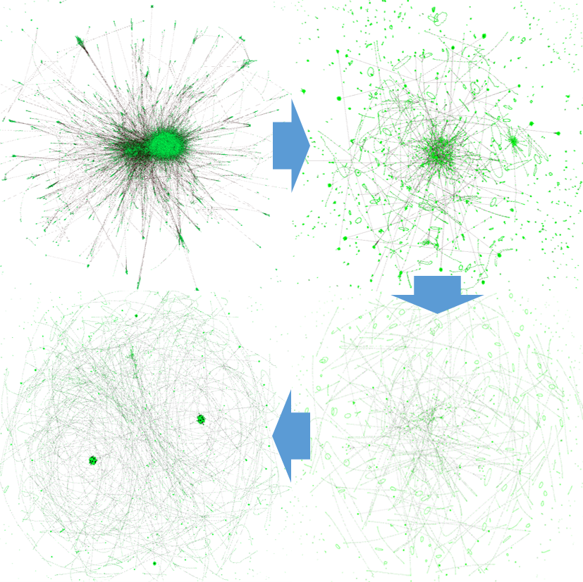}
\caption{Dynamics of a virtual goods trading network over time.}
\label{fig:change network}
\end{figure}

To analyze the temporal characteristics of virtual goods trading networks, we traced its dynamic variations for a year (see Fig.~\ref{fig:change network}). 

\begin{figure}[ht]
\centering
\begin{tabular}{c c}
\includegraphics[width=4cm]{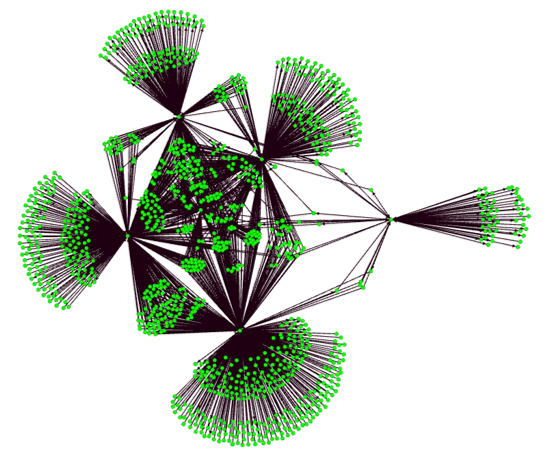} & 
\includegraphics[width=4cm]{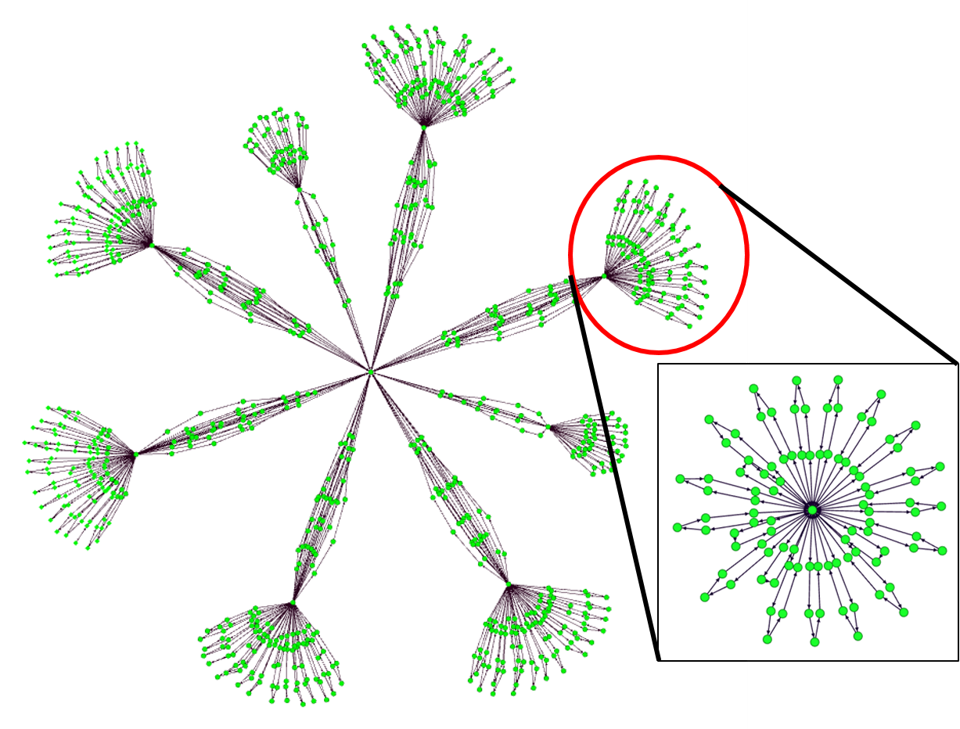} \\
\end{tabular}
\caption{Examples of oddly-shaped providers at old servers on Lineage.}
\label{fig:monopoly community}
\end{figure}

In the beginning, there exist many scattered local consumer communities due to that a large number of new users join and existing users leave the MMORPG; small-sized provider communities appear too. 
As time passes, the provider communities grew bigger and bigger. After a certain time interval, RMT consumption, which was boomed excessively in the beginning, decreases and a significant number of providers become extinct. After this period, a few extra-large provider communities, survivors, emerge. Finally, ``\emph{winner-takes-all}'' phenomena (with a small number of well-organized providers only) could be observed in the RMT market of online games.
In the process, we found several extremely large communities having an hierarchical structure (see Fig.~\ref{fig:monopoly community}). In the hierarchical network structure, nodes are divided into groups that are subdivided into smaller groups. According to Clauset, Moore and Newman's work ~\cite{clauset2008hierarchical} and Xu and Chen's work ~\cite{xu2005criminal}, a network of association between terrorists has a hierarchical structure. The hierarchical structure with many subgroups are effective to deliver the order from a central node and also minimizes the risk of network being collapsed even though one of subgroups are detected. The emergence of hierarchical structure indicates that the RMT provider communities become more systematically organized and at the same time, to minimize the risk of revealing.

\subsection{RMT network in other games}
\begin{figure}[ht]
\centering
\begin{tabular} {c c} \\
\includegraphics[width=3.5cm]{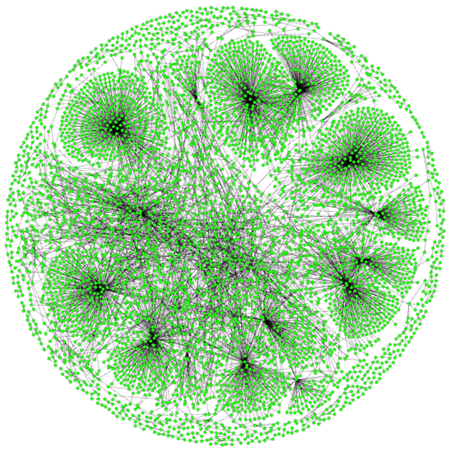} &
\includegraphics[width=3.5cm]{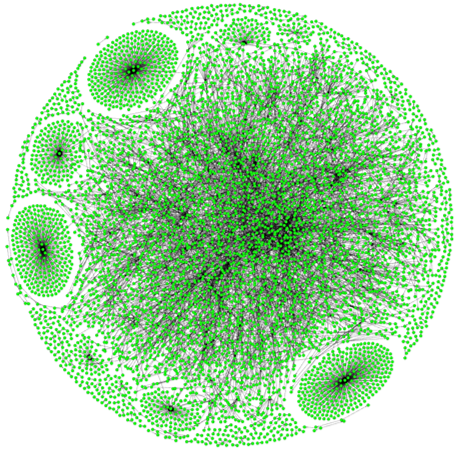} \\
(a) Aion & (b) Blade and Soul \\
\end{tabular}
\caption{Virtual goods trading networks in other MMORPGs}
\label{fig:aion network}
\end{figure}

To generalize our approach and findings from a game, Lineage, we applied our method to other similar genre games. Fig.~\ref{fig:aion network} is a visualization of the trading network in Aion and Blade and Soul, which are other famous MMORPGs of NCSOFT. Fig.~\ref{fig:aion network} shows that star-shaped communities, which are suspected as RMT providers, also are discovered in trading networks of the games. We will apply our methodology to other games' dataset in future work.

\section{Ethical considerations}
Before installation of a game client program, participants were asked to acknowledge a consent form under the end user license agreement (EULA) and domestic laws. This informed them that their data may be used in improving the quality of the installed game. Anonymous data samples were collected only for the purpose of conducting statistical analyses. Ethical compliance of our research went through an internal experimental review process at our organization before it launched (IRB and equivalents).

\section{Related Work}
\label{section:related works}

The study about RMT was originally initiated in the field of Economics. Huhh~\cite{huhh2008simple} viewed RMT in massively multi-online game in the perspective of network externality. Huhh pointed out that RMT has contributed to the commercial success of online games. 
Heeks~\cite{heeks2009understanding} analyzed GFG and RMT in a macroeconomic point. According to~\cite{heeks2009understanding}, RMT is essential to provide a channel to transfer assets between the virtual and real world. 

In this paper, we provide valuable information to prove the claims made in previous studies. For example, Heeks~\cite{heeks2009understanding} explained that RMT providers pursued economies of size and scale in the RMT market. We confirmed that this phenomenon was observed through our analysis of the virtual goods trading networks in the real game dataset.

Several studies analyzed the social network structures in online games because such games provide various social activities similar to the real world. Chung et al.~\cite{chung2014unveiling} and Son et al.~\cite{son2012analysis} studied various characteristics of social interaction networks in MMORPG. All of these studies could be used as a practical guide for understanding the social relationship between users.

Keegan et al.~\cite{keegan2010dark} particularly explored the behaviors of GFGs using a network analysis. They showed similarities between a GFG network and real world drug trafficking network. This is proof that analyzing a community network in an online game can assist in understanding a real world community structure. Woo et al.~\cite{woo2011can} analyzed the free money trade referring that people only give game money without getting any goods in return for GFG detection. Fujita et al.~\cite{fujita2011detecting} proposed a RMT detection method based on the observation that RMT traders tend to form dense relationships with each other. Although their works and our work had similar approaches in terms of analyzing trading network, we improved the RMT detection method by considering community network structures and member characteristics in a community while previous studies analyzed only trade activities.

Meanwhile, we refer to many previous studies for social network analysis. Mislove et al.~\cite{mislove2007measurement} is a representative study for a social network on the Internet. 
Yang et al.~\cite{yang2012analyzing} proposed a spammer detection model using network analysis in Twitter. Boshmaf et al.~\cite{boshmaf2011socialbot} studied the social BOT network in Facebook. 
Although these studies analyzed services on the Internet, their approaches and techniques were helpful in analyzing social networks in online games. We believe that our work can additionally contribute to understanding network structures of other Internet services.

\section{Conclusion}
\label{section:conclusions}
RMT could significantly influence online game services. The overproduction and popularity of RMT could have a negative impact on the virtual economy and cause severe damage to the reputation of the MMOPRG. On the other hand, RMT presents the potential revenue which online game company may get from In-app purchase. 
Therefore, it is critically important for game companies to estimate volumes of RMT and monitor the trend. In practice, however, it is very difficult to correctly distinguish RMT transactions from the millions of transactions. 

To address this problem, we propose to detect abnormal communities on a trade network formed from in-game transactions. We construct a trading network with virtual goods trading records, split hundreds communities, detect RMT groups by characterizing their unique network statistics and analyzing community members' in-game activities. We then filter out trades from RMT group, namely RMT provider to normal groups, namely RMT consumers and estimate the RMT volume based on filtered trades. To evaluate our RMT provider detection, we measured the rate of users who were banned by illegal activity (i.e. using game bot) belonging to each community type and found that RMT provider groups have more banning users than other types. 

For practical use, we performed the experiment using large-scale game logs from a popular online game---Lineage, one of the most popular MMORPGs. To demonstrate the effectiveness of the proposed method, we compared the estimation on RMT detected from in-game log with real RMT transaction records crawled from two major RMT websites in South Korea. Through the analysis on a large-scale and long-term log from a popular online game, we derived following findings.

Professional RMT providers typically form the star-shaped network, which is highly centralized with a high-degreed key node and lots of low-degrees nodes. This shape is for efficient sales of asset; the key node collect asset from members and sell it to other users. We also found that chained structures in the virtual goods trading network. This network is emerged to minimize the risk of revealing and the damage of banning. This network structure has never been found in previous works to our best knowledge. 
Additionally, we showed several interesting results about dynamics of RMT network via long-term analysis. As the online game become mature, the RMT market is monopolized by a few large RMT providers. Also, in old server, the RMT provider communities with the hierarchical structure appear for more systematic operation. We expect that this work is helpful to understand the underground economy in not only an online game but also real-world. As a part of future work, we plan to apply our methodology to other online games' dataset to generalize our findings.

Finally, our current analysis of RMT transactions only focused on the online game service. We believe the proposed method might work well for other black markets in other networks (e.g., Bitcoin transactions in the \emph{Deep Web}~\cite{He07:WEB}). We will also verify this as another part of future work.

\bibliographystyle{reference_format}
\bibliography{references}

\end{document}